\documentclass[aps,twocolumn,english,superscriptaddress,citeautoscript,preprintnumbers,amsmath,amssymb,floatfix,footinbib]{revtex4-1}
\usepackage{hyperref}
\hypersetup{colorlinks=true,linkcolor=red,citecolor=blue}
\usepackage{amsmath}
\usepackage{graphicx}
\usepackage{dcolumn}
\usepackage{float}

\begin{document}

\title{Lattice dynamics of topological Dirac semimetal LaAgSb$_{2}$ with charge density wave ordering}

\author{Ratnadwip Singha}

\thanks{These authors contributed equally to the work.}

\affiliation{Saha Institute of Nuclear Physics, HBNI, 1/AF Bidhannagar, Kolkata 700 064, India}

\author{Sudeshna Samanta}

\thanks{These authors contributed equally to the work.}

\affiliation{Center for High Pressure Science and Technology Advanced Research, Shanghai 201 203, China}

\affiliation{Micro-Nano System Center, School of Information Science and Technology, Fudan University, Shanghai 200 433, China}

\author{Tara Shankar Bhattacharya}

\affiliation{Department of Physics, Bose Institute, 93/1, Acharya Prafulla Chandra Road, Kolkata 700 009, India}

\author{Swastika Chatterjee}

\affiliation{Department of Earth Sciences, Indian Institute of Science Education and Research, Kolkata, 741 246, India}

\author{Shubhankar Roy}

\affiliation{Saha Institute of Nuclear Physics, HBNI, 1/AF Bidhannagar, Kolkata 700 064, India}

\affiliation{Vidyasagar Metropolitan College, 39, Sankar Ghosh Lane, Kolkata 700 006, India}

\author{Lin Wang}

\email{linwang@ysu.edu.cn}

\affiliation{Center for High Pressure Science and Technology Advanced Research, Shanghai 201 203, China}

\affiliation{Center for High Pressure Science (CHiPS), State Key Laboratory of Metastable Materials Science and Technology, Yanshan University,
Qinhuangdao, Hebei 066 004, China}

\author{Achintya Singha}

\email{achintya@jcbose.ac.in}

\affiliation{Department of Physics, Bose Institute, 93/1, Acharya Prafulla Chandra Road, Kolkata 700 009, India}

\author{Prabhat Mandal}

\email{prabhat.mandal@saha.ac.in}

\affiliation{Saha Institute of Nuclear Physics, HBNI, 1/AF Bidhannagar, Kolkata 700 064, India}

\date{\today}

\begin{abstract}

LaAgSb$_{2}$ is a rare material, which offers the opportunity to investigate the complex interplay between charge density wave (CDW) ordering and topology protected electronic band structure. As both of these phenomena are governed by the structural symmetries, a comprehensive study of the lattice dynamics is highly desirable. In this report, we present the results of temperature and pressure dependent Raman spectroscopy and x-ray diffraction in single crystalline LaAgSb$_{2}$. Our results confirm that Raman spectroscopy is a highly sensitive tool to probe CDW ordering phenomenon, particularly the low-temperature second CDW transition in LaAgSb$_{2}$, which appears as a very weak anomaly in most experiments. The crystal orientation-dependent measurements provide the evolution of Raman modes with crystallographic symmetries and can be further studied through group symmetry analysis. The low-temperature x-ray diffraction data show the emergence of structural modulations corresponding to the CDW instability. The combined high-pressure Raman spectroscopy and synchrotron x-ray diffraction reveal multiple structural phase transitions through lowering of crystalline symmetries, which are also expected to lead to electronic topological transitions.

\end{abstract}

\maketitle

\section{Introduction}

Charge density wave (CDW) ordering in a material is the periodic modulation of the electronic charge density, which spontaneously breaks the discrete translational symmetry of the lattice \cite{Gruner}. Thus, a new periodicity, determined by the nature of CDW (commensurate/incommensurate), is introduced. Such a quantum phase transition is further enhanced by Fermi surface nesting and the opening of either a full or partial gap at the Fermi surface strongly influences the charge conduction mechanism in the CDW state. The nature of gap (partial or full) at the Fermi surface is extremely sensitive to the dimensionality of the system. In one-dimension, it is easier to gap out the Fermi surface and induce a CDW insulating state. However, in two- or three-dimensional multiband systems, usually partial gap opening occurs below the CDW transition temperature ($T_{CDW}$). A partial gapping-out of the Fermi surface leads to strong reduction in the charge scattering, which enhances Landau-quasiparticle- like coherence in the CDW state. As a direct consequence, the system exhibits metallic behavior with high carrier mobility at low temperature below $T_{CDW}$. So, CDW state is mainly investigated in quasi-two-dimensional layered materials, as the nesting conditions are favorable in lower dimension.

LaAgSb$_{2}$ crystallizes in tetragonal layered structure, where La-Sb slabs are separated by Sb and Ag sheets \cite{Myers}. This material shows two distinct CDW transitions \cite{Myers,Song} along with unconventional transport properties such as large linear magnetoresistance, high carrier mobility, and opposite sign in the Hall and Seebeck coefficients \cite{Wang}. In fact, members of the $R$AgSb$_{2}$-family ($R$=rare earth elements) show a vast array of novel electronic and magnetic properties including the Kondo lattice state, crystalline electric field controlled magnetic anisotropy, metamagnetic transitions,  and ferromagnetic/antiferromagnetic ordering, etc \cite{Myers,Houshiar,Takeuchi,Myers2}. In recent years, LaAgSb$_{2}$ has drawn renewed interest as a topological Dirac semimetal \cite{Shi}. The angle-resolved photoemission spectroscopy measurements have confirmed the presence of Dirac cone formed by Sb 5$p_{x,y}$ orbitals along the $\Gamma$-M direction of the electronic band structure \cite{Shi}. Moreover, the same linearly dispersing bands are found to be responsible for the Fermi surface nesting and hence, the formation of CDW state in this material. Interestingly, the nature of the magnetoresistance also changes (from linear to quadratic dependence on magnetic field) above the CDW transition temperature \cite{Wang}.

It is well established that the underlying lattice in a material plays a crucial role for the occurrence of the CDW state. On the other hand, crystallographic symmetries protect the electronic band crossings and topologically non-trivial band structure. Therefore, a systematic investigation of the lattice dynamics in a material not only probes the phonon subsystem, but also provides fundamental information about the collective electronic excitations. In this paper, we have employed Raman spectroscopy and x-ray diffraction measurements to study the lattice dynamics of LaAgSb$_{2}$. With the aid of density functional theory calculations, we have combined experimental and theoretical approaches to assign different vibrational modes of the Raman spectra. In Raman modes, the clear signature of temperature- and pressure-induced crystallographic phase transitions motivated us to perform x-ray diffraction (XRD) measurements at low temperature and under compression. The low-temperature XRD spectra reveal structural modulations corresponding to the formation of CDW state. On the other hand, our synchrotron XRD results demonstrate pressure-induced tetragonal to lower symmetric orthorhombic phase transition in LaAgSb$_{2}$.

\section{Experimental and theoretical details}

The single crystals of LaAgSb$_{2}$ were grown by Sb-flux method \cite{Myers}. Elemental La (Alfa Aesar 99.9\%), Ag (Alfa Aesar 99.99\%), and Sb (Alfa Aesar 99.9999\%) in molar ratio 1:2:20, were taken in an alumina crucible. The crucible was then put in a quartz tube and sealed under dynamic vacuum. The quartz tube was heated to 1050$^{\circ}$C, kept at this temperature for 12 hrs, and slowly cooled (2$^{\circ}$C/h) to 670$^{\circ}$C. At this temperature, the crystals were separated from Sb-flux by centrifugation. The obtained crystals are rectangular, plate-like with crystallographic $c$-axis perpendicular to the plane. The transport measurements were done in a physical property measurement system (Quantum Design) using ac-transport option. Conducting silver paste and gold wires were used to make the electrical contacts in four-probe configuration.

The data for temperature-dependent Raman spectra were collected with a LABRAM HR 800 system, which is equipped with a spectrometer of 80 cm focal length, 1800 gr/mm grating, Peltier cooled CCD, and a liquid nitrogen cryostat. A 100$\times$ objective with NA 0.9 was used to focus the laser beam of 488 nm wavelength on the crystal. The spectra were acquired with a temperature step about 2 K near the CDW transition temperature and in 5 K interval for the rest of the measurements. Pressure dependence of Raman spectroscopy studies were performed using inVia Renishaw Raman spectrometer with an excitation laser wavelength 633 nm and grating 1800 gr/mm. The range of our interest was 50-200 cm$^{-1}$ and wavenumbers of Raman peaks were determined with resolution 1 cm$^{-1}$. The single crystal along with a ruby ball were loaded in a diamond-anvil-cell (DAC) with Rhenium gasket. Silicone oil was used as pressure transmitting medium (PTM). For pressure calibration, we used ruby R1 fluorescence line shift and standard ruby pressure scale. The Raman active modes were recorded with increasing pressure at room temperature.

Temperature dependence of powder XRD measurements on crushed single crystals of LaAgSb$_{2}$ were performed using the low-temperature stage of a high resolution Rigaku TTRAX III powder x-ray diffractometer with Cu K$_{\alpha}$ radiation. Data were collected at several temperatures over the scattering angle range 20$^{\circ}$-80$^{\circ}$. The synchrotron XRD studies were performed at Beijing Synchrotron Radiation Facility (BSRF) using a monochromatic beam with wavelength $\lambda$=0.6199 {\AA}. A symmetric DAC (culet size 300 $\mu$m) with Rhenium gasket and silicone oil as PTM were used. MAR-345 image plates were used for experiments to obtain the high quality powder diffraction patterns from the studied sample. The detector distance was calibrated with CeO$_{2}$ standard. The ring-type two-dimensional diffraction data were collected and one-dimensional intensity vs. diffraction angle (2$\theta$) patterns were generated using the Dioptas software program. The structural refinements were carried out using Jana2006 program.

All theoretical calculations have been performed using first principles density functional theory (DFT) as implemented in the plane-wave based \textit{Vienna ab initio Simulation Package} (VASP) \cite{Kresse:1993,Kresse:1996,Kresse:1996a,Kresse:1999}, within the choice of Generalized Gradient Approximation (GGA) for exchange-correlation functional as implemented in the Perdew-Burke-Ernzerhof (PBE-GGA) formulation \cite{Perdew:1996}. We have used the projector augmented wave (PAW) potentials and wave-functions were expanded in a plane wave basis with a kinetic energy cut-off of 900 eV. To ensure the convergence of electronic energies during the self-consistent run and the calculation of Hellmann-Feynman forces on atoms during structural optimization using conjugate gradient algorithm, we have used a convergence criterion of 1$\times$10$^{-8}$ eV and 1$\times$10$^{-3}$ eV/{\AA} respectively. The $\Gamma$-point phonon frequencies have been calculated using density functional perturbation theory (DFPT) \cite{Baroni:2001}, as implement in the VASP code.\\

\section{Results and Discussion}

\begin{figure}
\includegraphics[width=0.45\textwidth]{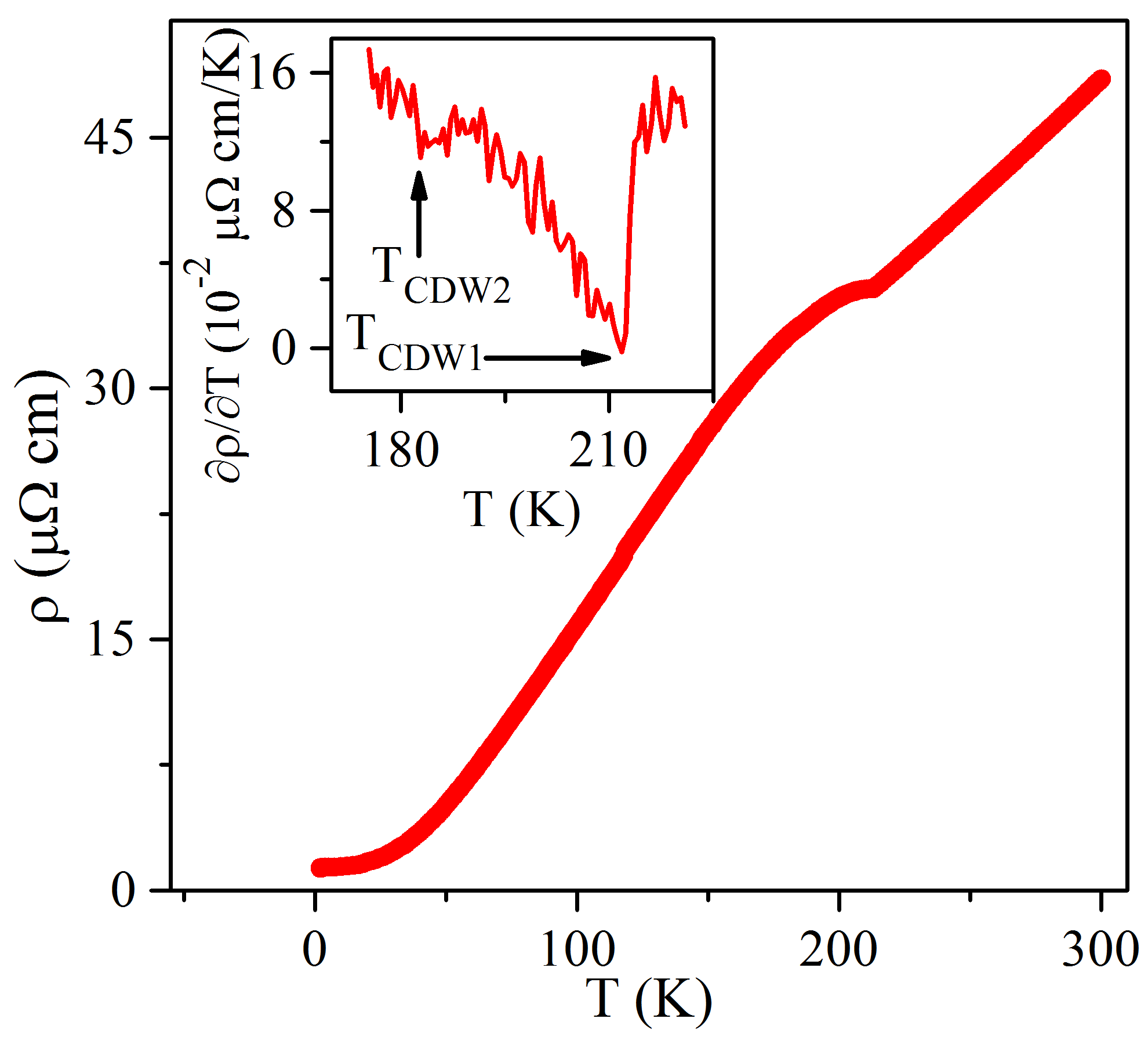}
\caption{(Color online) Temperature dependence of resistivity for LaAgSb$_{2}$ single crystal. Inset shows the two charge density wave transition temperatures.}
\end{figure}

\subsection{Temperature dependence of resistivity}

In Fig. 1, we have plotted the resistivity ($\rho$) for a typical LaAgSb$_{2}$ crystal as a function of temperature ($T$). The overall behavior is metallic in nature as the resistivity decreases monotonically with decrease in temperature except around 212 K, below which a hump-like feature appears. Similar anomaly in resistivity curve has also been observed in other systems at the CDW ordering temperature. In order to identify the CDW transition temperature, we have calculated the first-order derivative of the experimental data. Inset of Fig. 1 clearly shows two transition temperatures, $T_{CDW1}$=212 K and $T_{CDW2}$=184 K, which are consistent with the earlier reports \cite{Myers,Song}. The metallic behavior confirms the partial gap opening at the Fermi surface in CDW state.

\subsection{Raman spectra at ambient condition and crystal orientation dependent study}

\begin{figure*}
\includegraphics[width=0.8\textwidth]{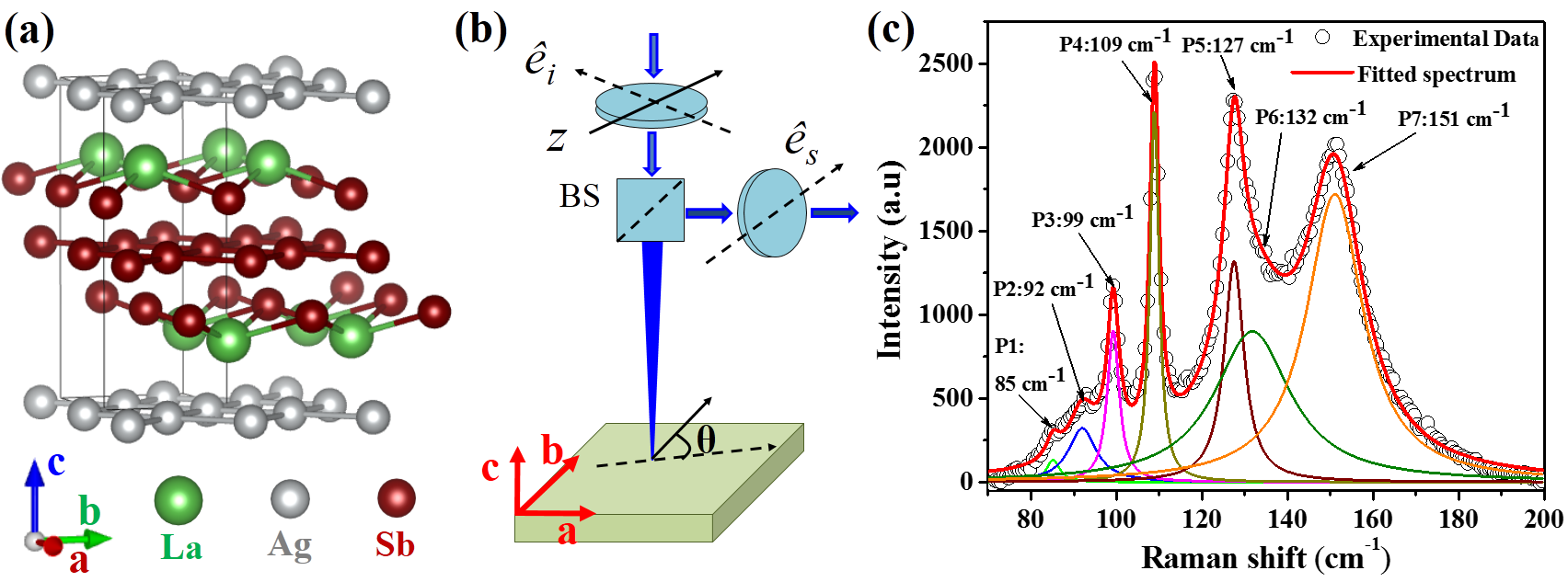}
\caption{(Color online) (a) Crystal structure of LaAgSb$_{2}$. (b) Schematic of the set-up for Raman scattering measurement. $\theta$ is the angle between the crystallographic $b$-axis and polarization vector of the incident light. (c) Raman spectra of LaAgSb$_{2}$ at room temperature showing seven Raman active modes.}
\end{figure*}

\begin{figure*}
\includegraphics[width=0.8\textwidth]{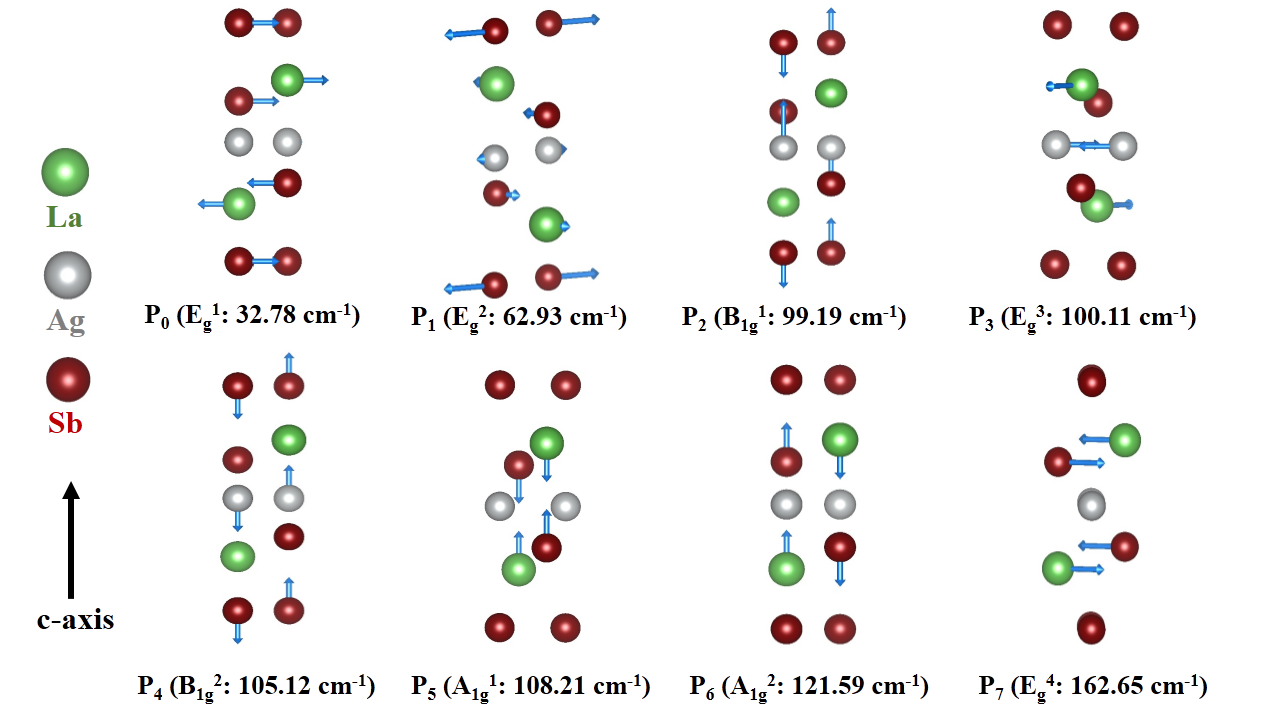}
\caption{(Color online) Theoretically calculated frequencies and vibration patterns corresponding to the Raman modes.}
\end{figure*}

\begin{figure*}
\includegraphics[width=0.7\textwidth]{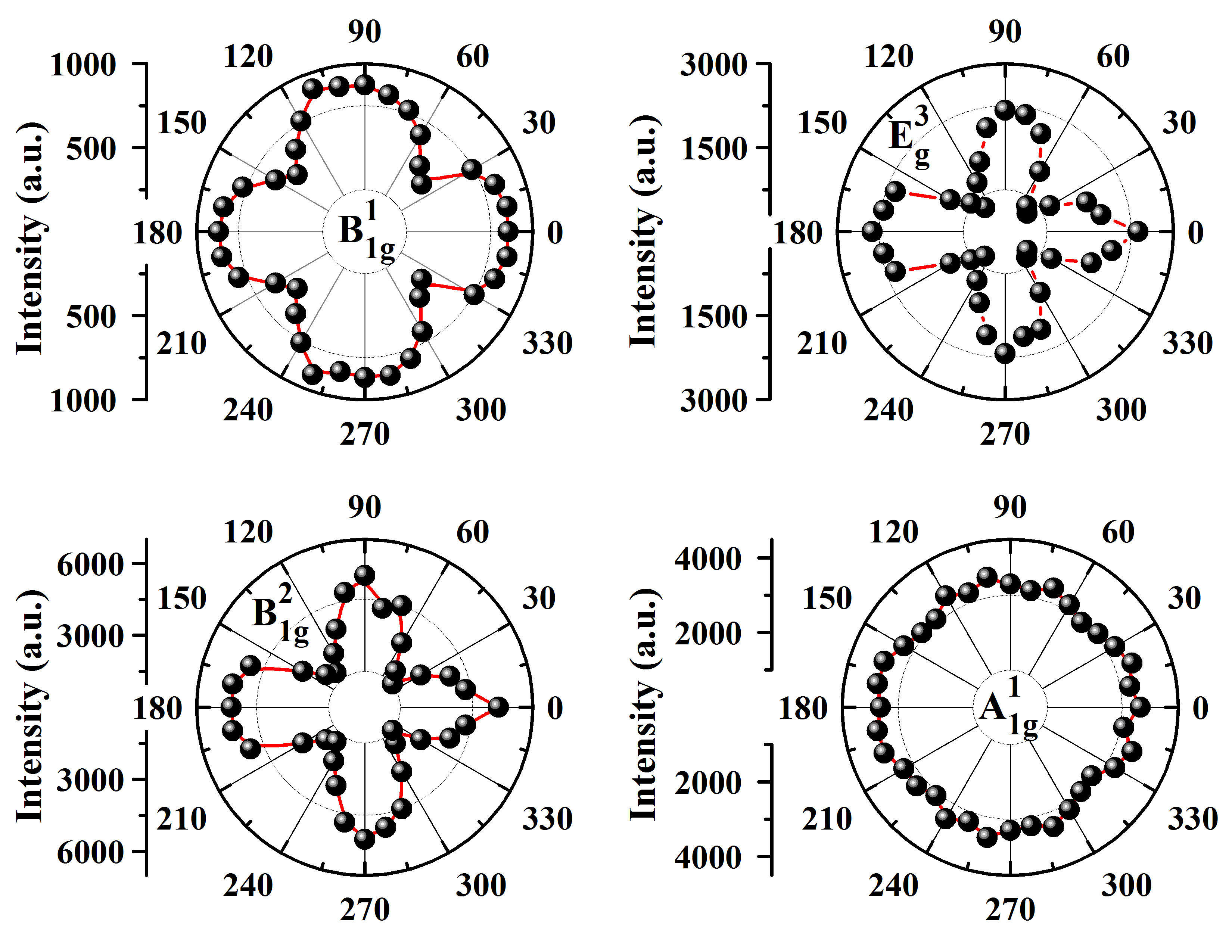}
\caption{(Color online) Angle dependence of intensity for $B_{1g}^{1}$, $E_{g}^{3}$, $B_{1g}^{2}$, and $A_{1g}^{1}$ Raman modes.}
\end{figure*}

Figures 2(a) and 2(b) illustrate the crystal structure of LaAgSb$_{2}$ and schematic of the used experimental set-up, respectively. The Raman spectroscopy measurements have been performed in backscattering geometry using a linearly polarized incident light with wave vector parallel to the crystallographic $c$-axis. The scattered light is also linearly polarized. For orientation dependent measurements (discussed in later section), the crystal is rotated about $c$-axis. Figure 2(c) shows the Raman spectrum of LaAgSb$_{2}$ at ambient condition. By fitting the experimental data, we are able to resolve seven Raman active modes (P1-P7). Among them, four namely P3 (99 cm$^{-1}$), P4 (109 cm$^{-1}$), P5 (127 cm$^{-1}$), and P7 (151 cm$^{-1}$) are most intense. In order to identify these modes, we have calculated the Raman dispersion spectra and phonon density of sates. Figure 3 illustrates theoretically determined atomic vibrational patterns, frequencies, and irreducible representations corresponding to all possible Raman active modes. From the detailed analysis, we have found that there are eight Raman active modes (2$A_{1g}$, 2$B_{1g}$, 4$E_{g}$) in LaAgSb$_{2}$. Among them, the first one (P0: $E_{g}^{1}$ 32.78 cm$^{-1}$) is well below the lower limit of our measured frequency range and hence could not be detected in present measurement set-up. The lowest frequency P0 ($E_{g}^{1}$) and intense mode P5 ($A_{1g}^{1}$) correspond to the out-of-phase vibration of La and Sb atoms. On the other hand, intense modes like P1 ($E_{g}^{2}$), P2 ($B_{1g}^{1}$), P3 ($E_{g}^{3}$), P4 ($B_{1g}^{2}$), and P7 ($E_{g}^{4}$) originate from 180$^{\circ}$ out-of-phase vibrations of individual or group of atoms in LaAgSb$_{2}$. For few modes, we have noticed a difference between experimentally obtained and theoretically calculated frequency. This is because the peak positions are very close to each other in the Raman spectra in addition to various approximations used for the theoretical calculations. In addition to the seven Raman active modes, we have detected three extremely weak kinks at 117, 139, and 162 cm$^{-1}$ in experimental Raman spectra. From group symmetry analysis, we conclude that these are not possible Raman modes for LaAgSb$_{2}$ and might appear from small contamination on the crystal surface. Although we confirm that excluding these weak peaks does not affect the peak position and FWHM of the intense modes significantly, we include their contribution during spectrum analysis in order to enhance the reliability of fitting (Appendix: Fig. 10).

Next, we have performed the crystal orientation-dependent Raman measurements that may provide fundamental information about the crystallographic symmetries. We start by fixing the polarization vector of the incident laser about the crystallographic $b$-axis. The crystal is then rotated slowly in small steps about the $c$-axis and the backscattered polarized Raman spectra are recorded [Fig. 2(b)]. This experimental set-up is equivalent to rotating the polarization vector in $ab$-plane. However, by rotating the crystal, we can avoid required additional optical pieces, which unintentionally modulate the intensity of different modes \cite{Ma}. In Fig. 4, we have plotted the intensity of the four Raman modes, which show prominent angle dependence. We have observed clear four-fold symmetric patterns for $B_{1g}^{1}$, $B_{1g}^{2}$, and $E_{g}^{3}$ modes, whereas the intensity of $A_{1g}^{1}$ mode shows weak angle dependence. With the exact knowledge of the Raman scattering tensors from group symmetry analysis, it is possible to understand these patterns.

\subsection{Temperature dependence of Raman modes}

\begin{figure*}
\includegraphics[width=0.8\textwidth]{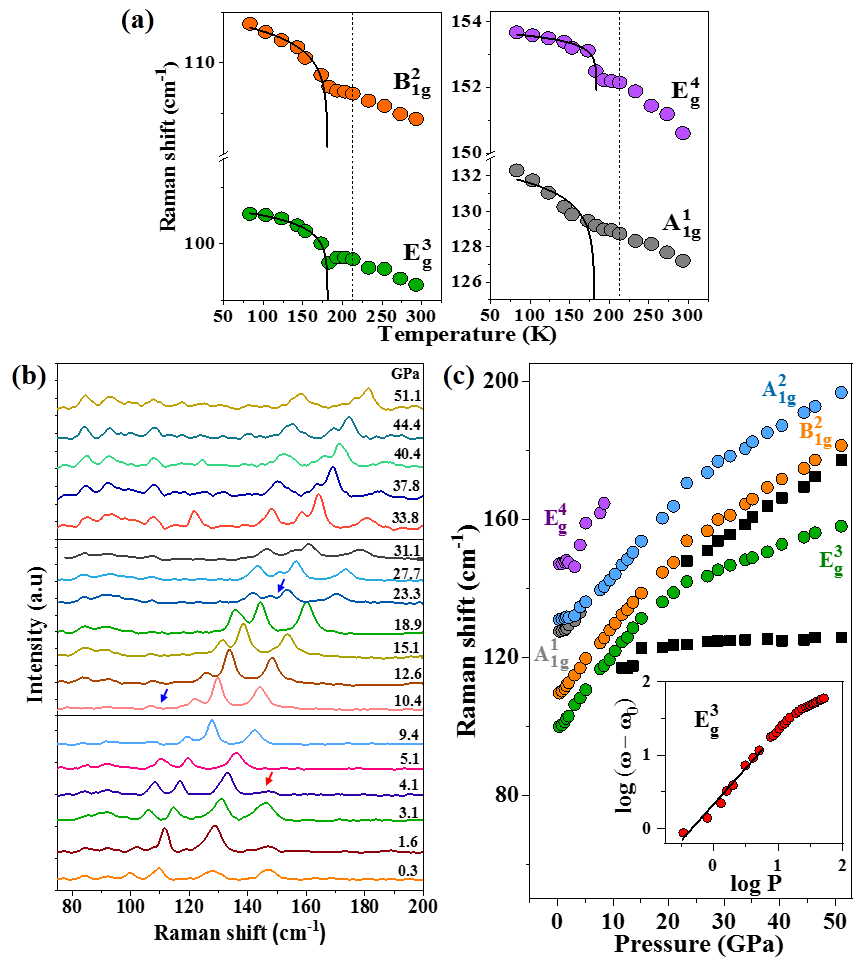}
\caption{(Color online) (a) Most intense Raman modes ($E_{g}^{3}$, $B_{1g}^{2}$, $A_{1g}^{1}$, and $E_{g}^{4}$) obey $\omega(T)$=$\omega^{\prime}$(1-$T/T_{C}$)$^{\beta}$ for $T<T_{C}$ (solid black curves). Vertical lines indicate the position of the first charge density wave transition. (b) Raman spectra of LaAgSb$_{2}$ at few representative pressures. The sudden disappearance of an active mode ($A_{1g}^{1}$) around 4.1 GPa and emergence of new peaks around 10.4 and 23.3 GPa are marked by red and blue arrows respectively. (c) The pressure dependencies of five Raman active mode frequencies. Two new modes (black filled squares) appear and continue to increase with high pressure up to 51 GPa. Inset shows log($\omega-\omega_{0}$) vs log$P$ plot for $E_{g}^{3}$ mode. The solid line shows the linear fit to the curve up to 4 GPa.}
\end{figure*}

To understand the subtle changes in the vibrational frequencies along with the formation and stability of the CDW phase, we have employed Raman spectroscopy both at low temperatures (50-300 K, ambient pressure) and under compression (0-51 GPa, 300 K). The obtained Raman spectra of LaAgSb$_{2}$ are shown in Appendix I at some representative temperatures (Appendix: Fig. 11). Figure 5(a) illustrates the temperature dependence of the Raman shift for four most intense modes ($E_{g}^{3}$, $B_{1g}^{2}$, $A_{1g}^{1}$, and $E_{g}^{4}$), which can be tracked down to the lowest temperature. All the modes demonstrate blue-shifts as the temperature is decreased, however, the modalities of the frequency shift are different due to different lattice anharmonicity coupled to them (Table I). Temperature dependencies of these modes are highly nonlinear in nature. All of them show a sharp increase in frequency below $\sim$180 K. No new peak has been observed to emerge during cooling down to the lowest temperature. Usually, the appearance of an amplitude mode for CDW ordering is the definite proof for collective excitation in a material. However, here, we have not observed such new mode. It is possible that the energy scale of the amplitude mode for LaAgSb$_{2}$ is below the accessible range of our experimental set-up.

LaAgSb$_{2}$ exhibits quasi-two-dimensionality in conductance and strongly coupled with its crystal structure. Watanabe \textit{et. al.} \cite{Watanabe} have reported the presence of two faint humps at 209 and 185 K, when the resistivity was measured with current along [001] direction in contrast to a single hump for [110] and [100] directions. Similar resistivity anomalies have also been observed for the present sample at $T_{CDW1}$ and $T_{CDW2}$ (Fig. 1). We note that the Raman mode frequencies show a sharp change near the second CDW transition temperature $T_{CDW2}$. Recently, a blue-shift along with sharp change in $E_{2g}^{1}$ mode have been reported at the incommensurate CDW transition temperature for TaSe$_{2}$ \cite{Hill}. Such behavior of $E_{2g}^{1}$ mode demonstrates the sensitivity of the in-plane Raman mode with distortion of the atomic coordinates in CDW phase. To understand the evolution of Raman modes with temperature below CDW transition in LaAgSb$_{2}$, we have fitted the phonon frequencies for $E_{g}^{3}$, $B_{1g}^{2}$, $A_{1g}^{1}$, and $E_{g}^{4}$ modes with well known power-law expression \cite{snow}, $\omega(T)$=$\omega^{\prime}$(1-$T/T_{c}$)$^{\beta}$, where $\omega^{\prime}$ is the phonon frequency at 0 K, $T_{c}$ and $\beta$ are the fitting parameters. Ideally, $\beta$ should be 0.5 according to the mean-field theory, but can differ significantly depending on various factors \cite{Shashidhar}. From the fitted curves [solid black lines in Fig. 5(a)], the transition temperature is found to be $T_{c}$=186 K, which is very close to $T_{CDW2}$ determined from $\rho$($T$) curve. Our results readily coincide with the temperature-dependent reflectivity study in LaAgSb$_{2}$, where a pronounced dip structure appears at 184 K in the frequency range 1000-200 cm$^{-1}$ \cite{Chen}. The authors noted this as the evidence for the second CDW transition with the formation of a charge gap at the Fermi level. This CDW ordering occurs along crystallographic $c$-axis with lattice modulation $\tau_{2}$$\sim$1/6(2$\pi$/$c$), as has been confirmed by x-ray scattering measurement \cite{Song} and is responsible for the weak anomaly in resistivity. The wave-vector of this second CDW locks at a low-temperature commensurate phase below 164 K with the value 1/6(2$\pi$/$c$). In comparison, we found that the Raman active modes in the present experiments also demonstrate high sensitivity to detect the hidden second CDW transition in LaAgSb$_{2}$, which was remained unexplored so far. On the other hand, for all modes, the peak positions show a weak change in slope near $T_{CDW1}$ [vertical lines in Fig. 5(a)]. It is also very interesting that at ambient conditions, the intensity ratio of two intense modes $A_{1g}^{1}$ (also $B_{1g}^{2}$) and $E_{g}^{4}$, i.e., $I_{A_{1g}^{1}}/I_{E_{g}^{4}}>$ 1 and it becomes unity around $T_{CDW1}$ [Fig. 2(c) and Appendix: Fig. 11]. As temperature decreases below $T_{CDW2}$, $E_{g}^{4}$ becomes the most intense mode, resulting $I_{A_{1g}^{1}}/I_{E_{g}^{4}}<$ 1. Signatures of both CDW transitions can also be detected from full width at half maximum (FWHM) data of these peaks (Appendix: Fig. 12).\\

\begin{center}
\begin{table*}
\caption{Anharmonicity contributions to intense Raman modes.}
 \begin{tabular}{|c c c c c c c c c|}
 \hline
  & Raman modes & $\omega$($P$, $T$) & ($\partial\omega$/$\partial P$)$_{T}$ & ($\partial\omega$/$\partial T$)$_{P}$ & $\gamma_{iT}$ & $\gamma_{iP}$ & $\eta$ & $\Sigma_{i}=$\\

  & & (cm$^{-1}$) & (cm$^{-1}$/GPa) & (cm$^{-1}$/K) & & & ($\gamma_{iT}$/$\gamma_{iP}$) & $\alpha_{c}$($\gamma_{iT}$-$\gamma_{iP}$)\\

  & & & $T$=300 K & $P$=1 atm & & & & $\times$10$^{-5}$ (K$^{-1}$) \\ [0.5ex]
 \hline\hline
  & $E_{g}^{3}$ & 99.88 & 2.079$\pm$0.019 & -0.00545$\pm$0.0004 & 1.73$\pm$0.02 & 1.21 & 1.42 & 2.34\\
  & $B_{1g}^{2}$ & 109.57 & 1.91489$\pm$0.023 & -0.00589$\pm$0.0002 & 1.54$\pm$0.01 & 1.19 & 1.29 & 1.57\\
 \hline
\end{tabular}
\end{table*}
\end{center}

\subsection{Pressure dependence of Raman modes}

Raman spectroscopy under pressure ($P$) may reveal important information on structural evolution of a compound. The Raman spectra of LaAgSb$_{2}$ at some selected pressures are displayed in Fig. 5(b). Out of seven modes, we could successfully track five intense Raman modes under compression. Few distinct changes have been observed, which are marked by arrows in Fig. 5(b). Around 4 GPa, the prominent $A_{1g}^{1}$ mode abruptly loses its intensity (red arrow) and does not reappear up to the highest applied pressure. We note that this mode corresponds to the vibration of La-Sb bonds along $c$-axis, where they are stacked between the gaps provided by alternating distorted two-dimensional nets of Sb-Sb and Ag-Ag atoms. With increasing pressure, as the interlayer spacing deceases rapidly, La-Sb layers are expected to be flattened and cease to vibrate anymore. Therefore, the compression-imposed lattice distortion and associated change in symmetry would cause $A_{1g}^{1}$ mode to disappear. Earlier electrical transport measurements on LaAgSb$_{2}$, for applied pressure well below 4 GPa, have shown that pressure stabilises metallic state and progressively suppresses $T_{CDW1}$ \cite{Bud'ko2,Torikachvili}. Our high pressure XRD results, to be discussed later on, suggest that around this pressure range, a new monoclinic phase starts to appear and both tetragonal and monoclinic phases coexist up to about 10 GPa. Further crystallographic transitions can be predicted as two new modes emerge around 10 and 23 GPa and survive up to 51 GPa as shown by solid black squares in Fig. 5(c). $E_{g}^{3}$, $B_{1g}^{2}$, $A_{1g}^{2}$, and these two new modes demonstrate hardening under compression. The phonon frequency curves also demonstrate change in slope around these critical pressure values ($\sim$4 and 10 GPa). In order to understand the pressure dependence of mode frequencies, we have plotted log($\omega-\omega_{0}$) vs. log$P$ for $E_{g}^{3}$ mode in Fig. 5(c) (inset), where $\omega_{0}$ is the frequency at ambient condition. We can readily identify three distinct pressure regions. In low-pressure region (up to 4 GPa), log($\omega-\omega_{0}$) vs. log$P$ is linear in nature with slope $\sim$1 [black solid line in Fig. 5(c):inset]. With increasing pressure, the slope decreases to $\sim$0.85 in the intermediate region and becomes $\sim$0.45 in the high pressure region.

In a material with positive thermal expansion coefficient, the vibrational frequencies are expected to decrease with increase in temperature. For LaAgSb$_{2}$, all of the modes exhibit negative temperature coefficient, ($\partial\omega$/$\partial T$)$_{P_{0}}$ $<$0. We have calculated the value of the coefficient for $E_{g}^{3}$ and $B_{1g}^{2}$ (Table I). These two modes are governed by mainly Ag-Ag, La-Sb, and Sb-Sb bonds and sensitive to any thermodynamic perturbation. The negative slope indicates temperature-induced bond expansion for these atoms in the lattice. Similarly, we have also calculated pressure coefficient of Raman modes ($\partial\omega$/$\partial P$)$_{T_{0}}$ and tabulated them (Table I). The positive pressure dependencies of both Raman modes confirm pressure-induced contraction of the bond distances. To calculate the isothermal Gr\"{u}neisen parameter ($\gamma_{iT}$), we have fitted the pressure dependent frequency shifts at room temperature using the relation,
\begin{equation}
\omega(P)=\omega(0)\left[1+P\frac{B_{0}'}{B_{0}}\right]^{\gamma_{iT}/B_{0}'}.
\end{equation}
We have used the reported values of bulk modulus $B_{0}$=74 GPa for LaAgSb$_{2}$ \cite{Bud'ko} and $B_{0}'$=4.0 \cite{Murnaghan}. The derived values of $\gamma_{iT}$ for $E_{g}^{3}$ and $B_{1g}^{2}$ modes are 1.73 and 1.54, respectively. On the other hand, with thermal expansion of the volume, the vibrational frequencies shift accordingly and can be described by isobaric Gr\"{u}neisen parameter ($\gamma_{iP}$). Using volumetric thermal expansion coefficient $\alpha_{c}$$\approx$4.5$\times$10$^{-5}$ K$^{-1}$ \cite{Bud'ko}, we have estimated values of $\gamma_{iP}$ as 1.21 and 1.19 for $E_{g}^{3}$ and $B_{1g}^{2}$ modes, respectively. Note that the calculated values of $\gamma_{iT}$ are higher than $\gamma_{iP}$ at ambient conditions. Hence, there is a crystal anharmonicity, quantified by an intrinsic anharmonic parameter $\Sigma_{i}$=$\alpha_{c}$($\gamma_{iT}$-$\gamma_{iP}$), which is positive. It represents the changes in electronic structure of the chemical bonds in LaAgSb$_{2}$ lattice. The calculated $\Sigma_{i}$ value is higher for $E_{g}^{3}$ mode as it originates from the planer vibrations of Ag-Ag and La-Sb bonds.\\

\begin{figure*}
\includegraphics[width=0.9\textwidth]{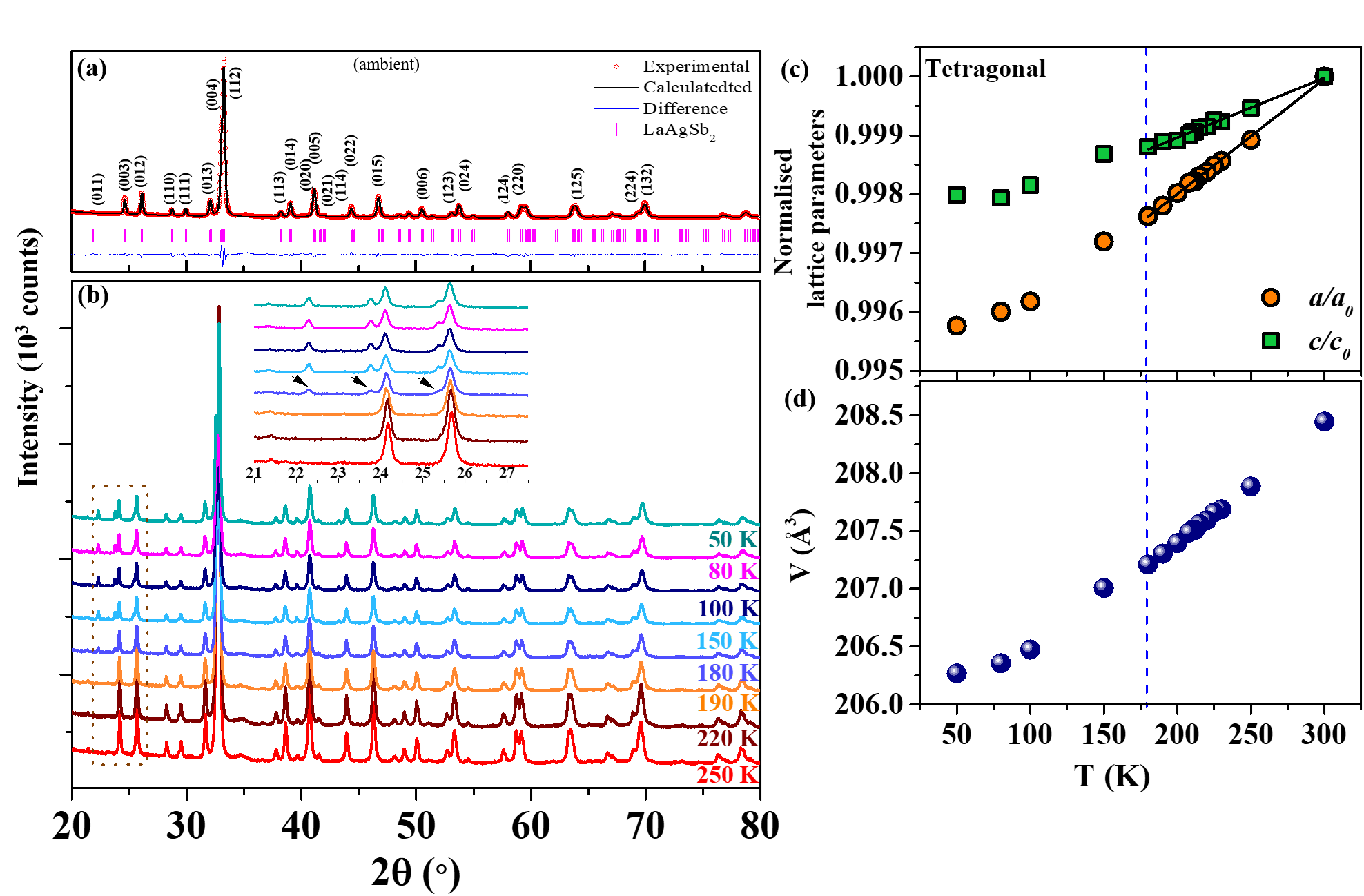}
\caption{(Color online) (a) XRD pattern of LaAgSb$_{2}$ at ambient condition, refined with $P4/nmm$ tetragonal structure: $Z$=2, $a$=$b$=4.416(2) {\AA} and $c$=10.891(1) {\AA}, $V$=212.386 {\AA}$^{3}$. (b) Temperature-dependent XRD patterns in the range 300-50 K showing clear phase transition around 180 K. The emergence of three new peaks are shown in the inset. (c) Temperature-dependent normalized lattice constants for tetragonal phase, where solid lines show their linear dependencies with temperature. (d) The change in volume with temperature for tetragonal phase.}
\end{figure*}

\subsection{Structural evolution at low-temperatures}

The typical powder x-ray diffraction pattern of LaAgSb$_{2}$ is shown in Fig. 6(a). All the peaks can be indexed well using the ZrCuSiAs-type tetragonal structure (space group: $P4/nmm$) with lattice parameters $a$=$b$=4.416(2) {\AA} and $c$=10.891(1) {\AA} \cite{Myers}. Thus XRD data confirm single phase of the material.

Low temperature XRD measurements for LaAgSb$_{2}$ reveal a structural phase transition around 180 K upon cooling as shown in Fig. 6(b) (inset: enlarged view of the region of interest). The results show that while the widths of (004) and (112) peaks are nearly unchanged upon cooling, the widths of (012) and (003) peaks  begin to increase significantly at about 180 K. Below this temperature, x-ray scattering study has reported the emergence of new satellite peaks with modulation wave vector 1/6(2$\pi$/$c$) along the $c$-axis \cite{Song}. Such modulation explains the resistivity anomaly at $T_{CDW2}$, which is more prominent in out-of-plane measurements. The emergence of three new peaks confirms that a new phase starts to appear below $T_{CDW2}$. LaAgSb$_{2}$ also exhibits another lattice modulation along $a$-axis with a wave vector 0.026(2$\pi$/$a$) corresponding to the first CDW transition at $T_{CDW1}$ \cite{Song}. However, the order parameter extracted from both of these modulations, shows much sharper transition at $T_{CDW2}$ in comparison to a weak anomaly at $T_{CDW1}$ \cite{Song}. Therefore, the lattice modulation must be much more pronounced near $T_{CDW2}$. This also explains the higher sensitivity of the second CDW ordering to Raman spectroscopy, which essentially captures the subtle changes in lattice. The normalized lattice parameters and unit cell volume for tetragonal phase are plotted in Figs. 6(c) and 6(d), respectively. Both lattice parameters deviate from linear dependencies [solid lines in Fig. 6(c)] below 180 K. It has been found earlier that the interatomic distances of La-Ag and Sb1-Sb2 (Sb1 in Sb layer and Sb2 in La-Sb slab) reach minima at 200 K, whereas La-Sb1 and La-Sb2 interatomic distances attain maxima at the same temperature \cite{Gondek}. The anisotropic evolution of lattice parameters starts at $T$$<$180 K and hence, the low-temperature structural transition is possibly linked to the CDW instability. The softening of the acoustic phonon modes also triggers such instability as recorded by the ultrafast pump probe spectroscopy \cite{Chen}.

\subsection{Structural stability under high pressure}

\begin{figure*}
\includegraphics[width=0.7\textwidth]{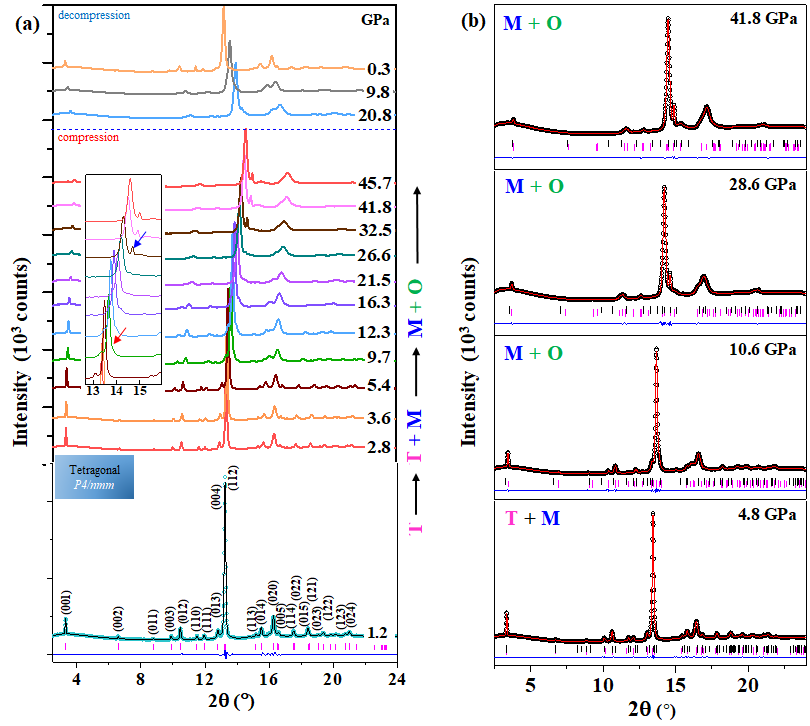}
\caption{(Color online) (a) (Bottom panel): Observed (open light blue circles), calculated (solid black line), and difference (blue line) profile obtained after profile refinement for LaAgSb$_{2}$ at 1.2 GPa at 300 K having tetragonal (space group $P4/nmm$) phase. (Middle panel): Pressure evolution of powder-synchrotron x-ray diffraction patterns at 300 K for selected pressure values under compression. Inset provides an enlarged representation within diffraction angle 13-15$^{\circ}$ for clarity. The arrows show the splitting (red arrow) and emergence of a new peak (blue arrow) around 10 and 26 GPa respectively. A reversible nature of phase transition has been recorded during decompression. `T', `M', and `O' indicate tetragonal, monoclinic, and orthorhombic phases respectively. (Top panel): The diffraction patterns at few representative pressure values during decompression. (b) The patterns for selective pressure 4.8, 10.6, 28.6, and 41.8 GPa are refined using combinations of tetragonal ($P4/nmm$), orthorhombic ($Cmma$), and monoclinic ($P112/n$) phases. The lower and upper vertical tick markers represent the Bragg positions for first and second coexisting phases, respectively.}
\end{figure*}

Figure 7 illustrates the results of the powder synchrotron XRD experiments under high pressure. We have obtained a high-quality refinement at low-pressure around 1.2 GPa without considering preferred orientation as shown in Fig. 7(a): bottom panel. We have also successfully distinguished the high pressure phases without considering any preferred orientation in order to reduce the refinement parameters. Being a layered compound, the inhomogeneity in LaAgSb$_{2}$ induced by the PTM, can not be ruled out.

\begin{figure*}
\includegraphics[width=0.9\textwidth]{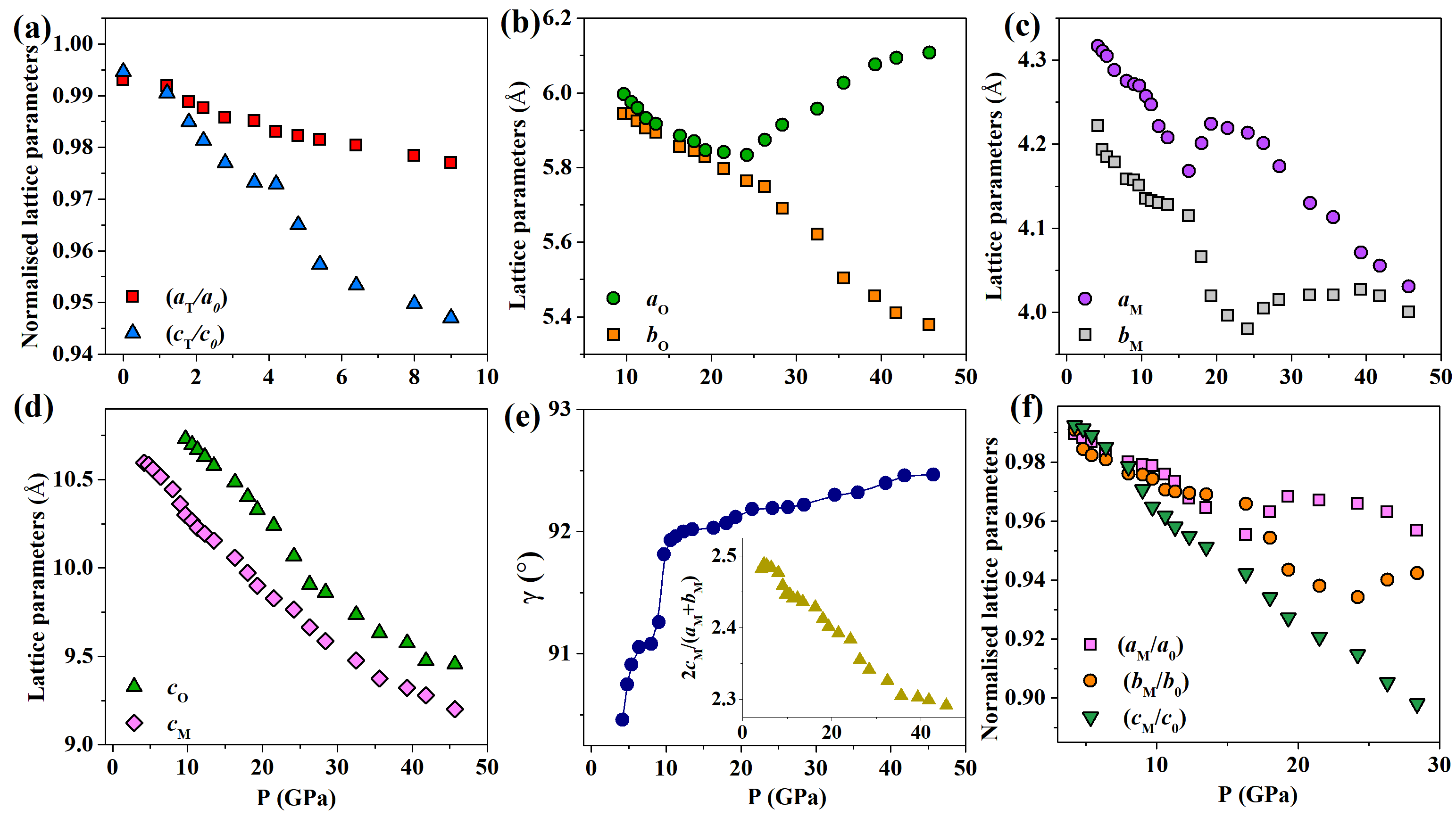}
\caption{(Color online) (a)-(d) Pressure evolution of structural parameters corresponding to three different crystallographic phases. The suffixes `T', `M', and `O' represent tetragonal, monoclinic, and orthorhombic phases, respectively. (e) Pressure dependence of the monoclinic angle $\gamma$, showing an abrupt increment around 10 GPa at the emergence of orthorhombic phase. Inset illustrates the change in inter-layer spacing in monoclinic phase compared to other crystallographic directions. (f) Normalized lattice parameters of the monoclinic phase with respect to the orthorhombic phase.}
\end{figure*}

Detailed analysis of diffraction pattern reveals that LaAgSb$_{2}$ retains its tetragonal symmetry below 4 GPa, which is consistent with the linear pressure dependencies of the intense Raman modes in the low-pressure regime [inset of Fig. 5(c)]. Figure 7(a) (middle panel) depicts the powder diffraction data with increasing pressure up to 45.7 GPa. Upon compression, significant broadening/splitting of (00$l$) peaks ($l$=1, 2, 4) has been observed. Though a very clear splitting of (220)$_{T}$ peak of tetragonal phase has not been observed, which is an unambiguous signature of the emergence of an orthorhombic phase, tetragonal phase alone fails to match the profiles. Such additional broadening may arise either from lowering of symmetry or due to the coexisting phases, as has also been reported for isostructural compounds. Hence, we have processed the refinement with coexisting tetragonal and orthorhombic ($Cmma$) structures up to 9 GPa. With increasing pressure, the extracted lattice parameters along with the cell volume decrease in a systematic fashion. Further increase in pressure gives rise to a shoulder peak next to (112)$_{T}$ [red arrow in Fig. 6(a)] around 10 GPa. As a result, the refinement with same tetragonal and orthorhombic structures appears to be insufficient. In addition, we can not index a signature peak (110)$_{T}$ using either tetragonal or orthorhombic structure above 4 GPa. Therefore, with increasing pressure, the same combination fails to explain the subtle structural changes with progressive deterioration of the refinement quality.

Critical inspection of diffraction pattern at 4 GPa  shows broadening in some peaks without any additional peak/splitting. Therefore, we have carried out refinements with various possibilities to identify the correct space group symmetry corresponding to the structural changes in LaAgSb$_{2}$ under high pressure. We have found that a distorted monoclinic structure with space group $P112/n$ coexists with tetragonal phase at low pressure and can be used to successfully index all the peaks. The volume of such monoclinic phase is smaller than that for the tetragonal phase by 3.3\% at 4.2 GPa. Such distortion in the crystal structure indeed supports the sudden disappearance of a Raman mode ($A_{1g}^{1}$: 127 cm$^{-1}$) at that pressure. An abrupt structural distortion with similar monoclinic phase has been probed by the neutron-scattering studies in isostructural LaOFeAs and explains its resistivity anomalies at low temperature \cite{Cruz}. For LaAgSb$_{2}$, there is a narrow pressure range (4-9 GPa), where tetragonal and monoclinic phases coexist and profile matches very well with experimental data [Fig. 7(b): bottom panel]. The signature peak (110)$_{T}$ can now be successfully indexed by the combination of (1$\bar{1}$0)$_{M}$ and (110)$_{M}$ of monoclinic phase. The observation of phase coexistence is also supported by the survival of the intense Raman mode $E_{g}^{4}$ up to 10 GPa along with the other hard modes ($E_{g}^{3}$, $B_{1g}^{2}$, and $A_{1g}^{2}$). For clarity, we have plotted the normalized lattice parameters ($a$/$a_{0}$ and $c$/$c_{0}$) for tetragonal phase in Fig. 8(a). Around 4 GPa, $c$-axis shows a sudden reduction and remains highly compressible compared to $a$-axis. This result supports our previous conjecture on anisotropic compression of the initial tetragonal phase, where the distortion-induced monoclinic phase emerges. The cumulative effect of lattice distortion and anisotropic lattice compression are evident in the non-linear pressure dependencies of the Raman active modes [Fig. 5(c): inset].

\begin{figure}
\includegraphics[width=0.45\textwidth]{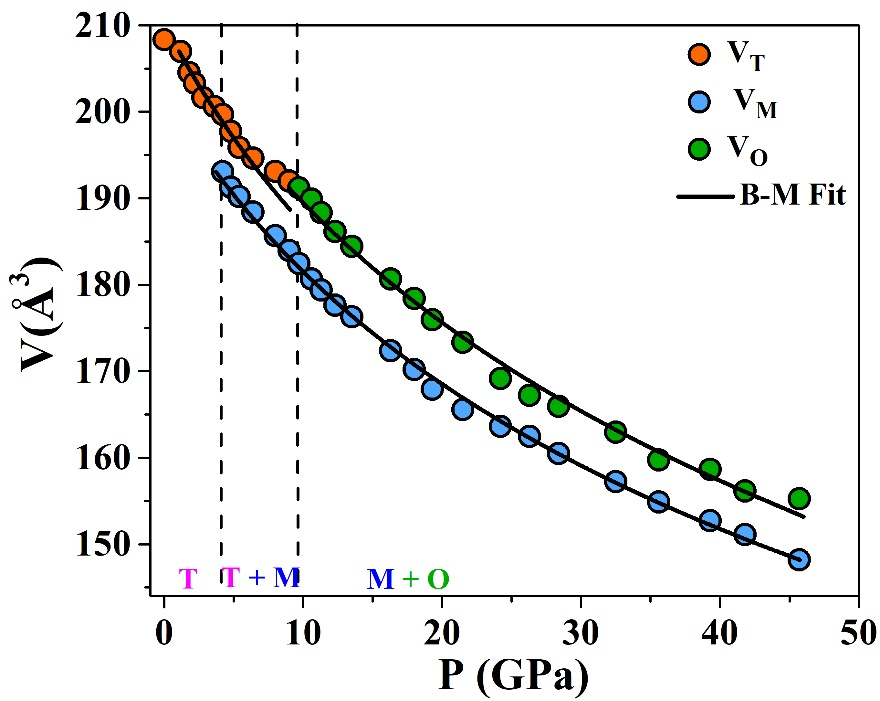}
\caption{(Color online) Volume corresponding to the three observed phases in LaAgSb$_{2}$, as a function of pressure. The black solid lines represent the Birch-Murnaghan equation of state fit for all three phases. The coexisting phases are separated by dashed lines.}
\end{figure}

To extract the pressure evolution of structural parameters, detailed LeBail refinements of x-ray diffraction patterns have been performed. At a critical pressure of about 9.7 GPa, powder diffraction pattern of LaAgSb$_{2}$ exhibits significant changes. At this pressure, the diffraction spectra can not be refined anymore using tetragonal and the monoclinic phases. We have carefully analyzed the (00$l$) peak profiles as $c$-axis is highly compressible compared to other two axes. The lower symmetric monoclinic phase appears less compressible than the tetragonal one. Between 4-10 GPa, the interlayer $c$-axis is more compressible than the $a$-axis in tetragonal phase because the linear compressibility of $c$-axis ($\varepsilon_{0}^{c}$=8.1 GPa$^{-1}$) is significantly higher than that of $a$-axis ($\varepsilon_{0}^{a}$=2.9 GPa$^{-1}$) \cite{Bud'ko2}. In comparison, the linear compressibility of $c$-axis for the monoclinic phase appears to be much lower than its tetragonal counterpart. Therefore, both tetragonal and monoclinic phases experience anisotropic compression, but the compression is more anisotropic for the tetragonal phase.

The tetragonal phase compels to transform into an orthorhombic phase above 10 GPa without breaking or reforming any bond in the crystal structure. We observe that an inclusion of orthorhombic phase with slightly larger lattice parameters [$Cmma$: $a$=5.9432(2) {\AA}, $b$=5.9932(2) {\AA} and $c$=10.7291(1) {\AA}] can significantly improve the refinement. The fits of the experimental profiles are satisfactory as depicted in Fig. 7(b). Therefore, two-phase refinement with monoclinic and orthorhombic phases has been proved successful for indexing all the observed diffraction peaks. The shoulder peak emerged at 10.6 GPa next to (00$l$)$_{T}$ peak, a prominent signature of the orthorhombic phase, can be indexed as (202)$_{O}$. The emergence of a new Raman mode around 10.4 GPa [Fig. 5(c)] further supports our observation. Both of these phases coexist up to the highest applied pressure $\sim$45 GPa) [Fig. 7(b):  top three panels]. The present results closely match with the earlier observations for isostructural SrFeAsF and CaFeAsF compounds, where orthorhombic and monoclinic phases coexist over a large range of pressure \cite{Mishra}.

For better understanding, we have plotted pressure dependence of structural parameters of LaAgSb$_{2}$ in Figs. 8(b)-(e). For easy comparison, the orthorhombic lattice parameters have been scaled to $a$=$a_{O}$/$\sqrt{2}$, $b$=$b_{O}$/$\sqrt{2}$, and $c$=$c_{O}$, where volume $V$=$V_{O}$/2; and $a$=$a_{M}$, $b$=$b_{M}$, and $c$=$c_{M}$ for monoclinic phase. Figures 8(b) and 8(c) demonstrate the anomalous variation of lattice constants for orthorhombic and monoclinic phases, respectively. For monoclinic phase, with increasing pressure above 17 GPa, the lattice parameter $a$ shows a sudden increase, whereas the parameter $b$ reduces rapidly up to 26 GPa. The opposite pressure dependence of $a$ and $b$ has also been observed for orthorhombic phase beyond 26 GPa. In contrast, $c$ axis shows monotonous pressure dependence for both of these phases [Fig. 8(d)]. Our results manifest that the pressure effect on lattice parameters is highly anisotropic in nature. We note that for an isostructural material LaMnPO, a tetragonal to orthorhombic phase transition has been reported followed by the orthorhombic to collapsed-orthorhombic transformation at 16 and 30 GPa, respectively \cite{Simonson}. This sudden collapse in all lattice parameters for LaMnPO, in absence of apparent change in crystal symmetry, indicates the extreme flattening of Mn-P layers under compression. Similarly, for LaAgSb$_{2}$, the discontinuity in monoclinic angle ($\gamma$) indicates the first-order structural phase transition at 10 GPa [Fig. 8(e)]. Moreover, a sudden contraction in $c_{O}$ [along (001)] around 26 GPa signifies a huge reduction in interlayer spacing compared to other directions [inset Fig. 8(e)]. The relative changes in the lattice paramters of two high-pressure phases are plotted in Fig. 8(f).

Finally, Fig. 9 shows the evolution of volumes for different phases with applied pressure. The Birch-Murnaghan equation of state fitting gives us the estimated bulk moduli for three phases as $B_{0}^{T}$=67.5 (1.7), $B_{0}^{M}$=72.6 (1.4), and $B_{0}^{O}$=76.3 (0.7) GPa for tetragonal, monoclinic, and orthorhombic phases, respectively, where we have kept $B_{0}'$=4.0 \cite{Murnaghan}. The value of bulk modulus of the tetragonal phase matches very well with the earlier report \cite{Bud'ko2}.

The present experiments clearly demonstrate that the application of high pressure first distorts the most compressible tetragonal phase of LaAgSb$_{2}$. Then, a pressure-induced distorted monoclinic phase appears and coexists with the parent phase with slightly higher compressibility. This monoclinic phase continues to sustain under compression and reduces its volume monotonically with pressure (Fig. 9). Above 10 GPa, the extreme compression causes another phase transition via orthorhombic distortion. The analysis performed above 26 GPa shows that the newly emerged peak corresponds to the monoclinic phase reflection (112)$_{M}$. In addition, a new Raman mode has appeared next to $B_{1g}^{2}$ mode and is directly related to the vibration of Ag-Ag and Sb-Sb along $c$-axis. Therefore, the newly emerged peaks in x-ray diffraction and vibrational spectroscopy are consistent with certain structural changes under compression. We have successfully detected the presence of monoclinic phase, considering reflections (1$\bar{1}$2)$_{M}$ and (112)$_{M}$ along with two orthorhombic reflections (020)$_{O}$ and (022)$_{O}$ around the strongest intensity peak. The newly emerged peak (112)$_{M}$ gradually increases its intensity with pressure as observed earlier for one of the peaks in isostructural layered ZrSiS \cite{Singha}. Such an increase in the intensity of a particular peak occurs due to the subtle change in $c/a$ ratio and has  been clearly observed for monoclinic phase [Figs. 8(c) and 8(f)].\\

\section{Conclusions}

To summarize, we have studied the lattice dynamics of topological Dirac semimetal LaAgSb$_{2}$ using Raman spectroscopy and x-ray diffraction measurements. We confirm that Raman spectroscopy is a highly effective tool in identifying CDW state in a material. In particular, this technique can be employed to find weak CDW transitions, which remain hidden in most experiments. The crystal orientation dependent measurements in LaAgSb$_{2}$ provide fundamental information about structural symmetries. From theoretical analysis, we have determined the lattice vibration spectra corresponding to all the Raman active modes. Our low-temperature x-ray diffraction data reveal signatures of lattice modulation corresponding to the CDW instability. High pressure Raman spectroscopy and synchrotron x-ray diffraction spectra show several pressure-induced structural phase transitions through lowering of crystallographic symmetries. As these symmetries play a fundamental role in protecting topologically non-trivial electronic band structure, we also predict pressure-induced electronic topological transitions in this material.

\section{Acknowledgement}

We thank Prof. Pranab Choudhury for his help and discussions during the calculations of irreducible representations of the Raman modes.

\section{Appendix}

\begin{center}
\textbf{I. Spectrum analysis using ten Raman modes.}
\end{center}

In Fig. 10 we have fitted the room temperature Raman spectra of LaAgSb$_{2}$ using ten possible modes (seven Raman active modes and three weak peaks due to surface contamination). By comparing these fitting results with Fig. 2(c), we confirm that the peak positions and FWHM of the intense modes are almost identical in both analysis.

\begin{figure}
\includegraphics[width=0.4\textwidth]{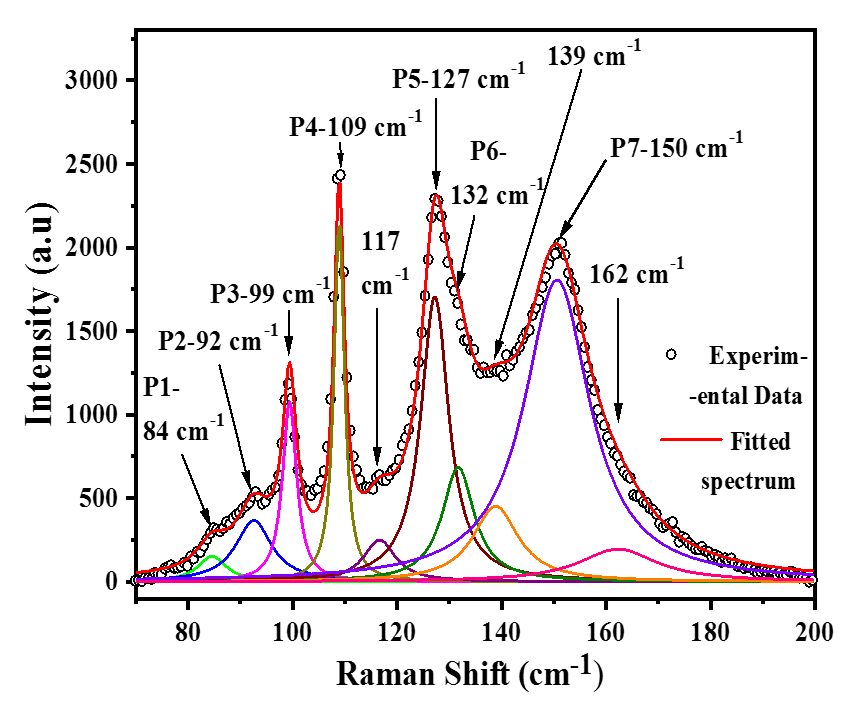}
\caption{(Color online) Analysis of room temperature Raman spectra for LaAgSb$_{2}$ using ten modes.}
\end{figure}

\begin{center}
\textbf{II. Raman spectra at different temperatures.}
\end{center}

Figure 11 shows the Raman spectra for LaAgSb$_{2}$ at some representative temperatures.

\begin{figure}
\includegraphics[width=0.4\textwidth]{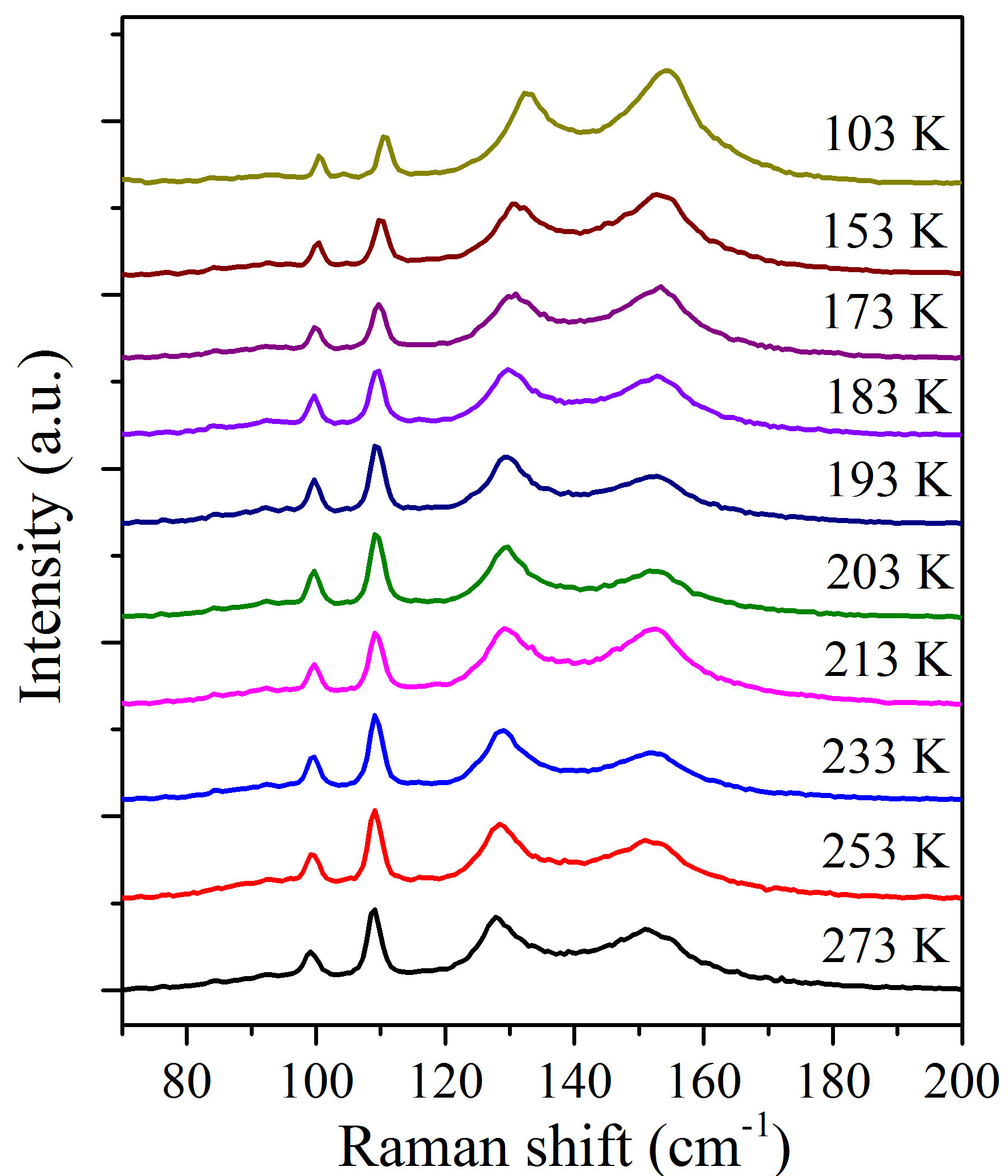}
\caption{(Color online) Raman spectra for LaAgSb$_{2}$ at some representative temperatures.}
\end{figure}

\begin{center}
\textbf{III. Full width at half maximum for most intense Raman modes.}
\end{center}

Figure 12 shows the temperature dependence of the full width at half maximum (FWHM) for most intense Raman modes. The vertical lines represent the transition temperature for two charge density wave orderings.

\begin{figure}[H]
\includegraphics[width=0.5\textwidth]{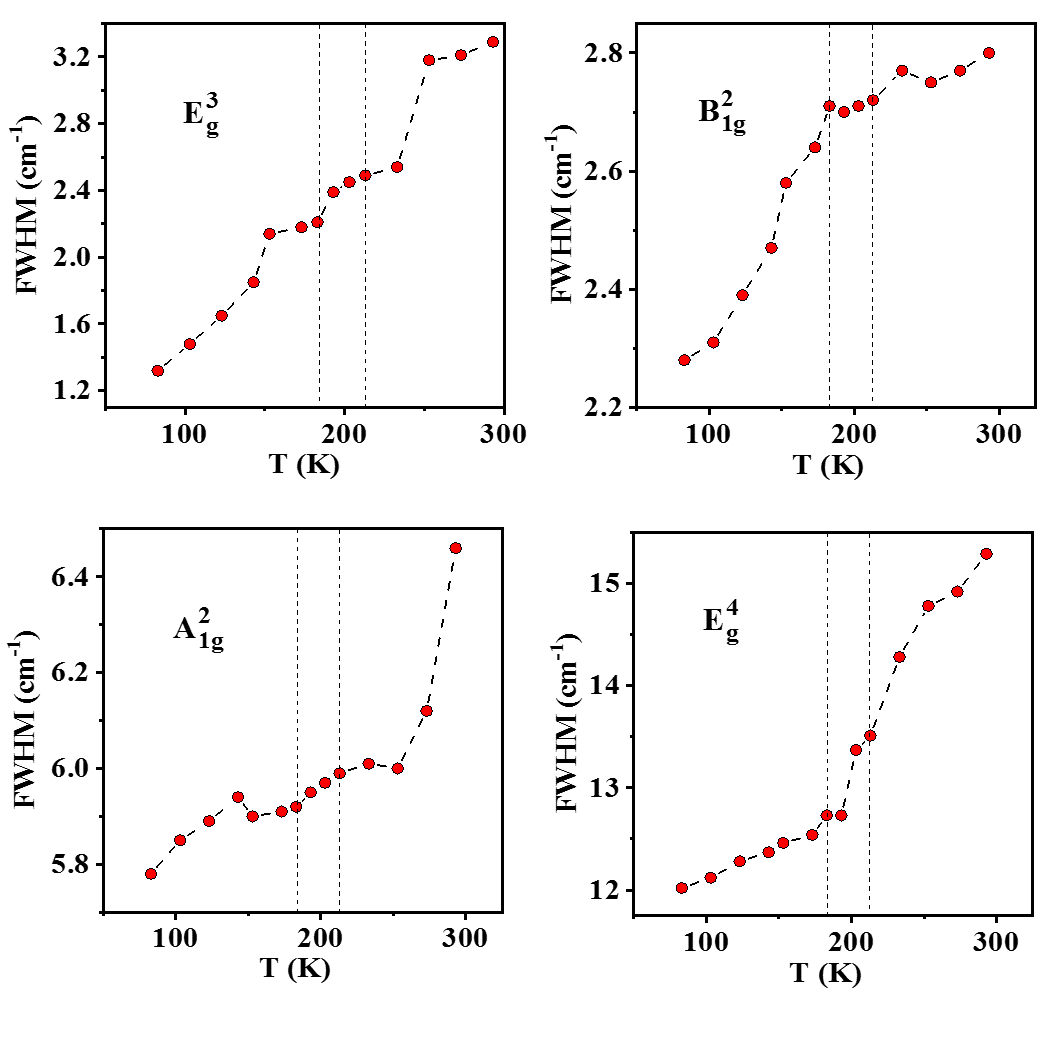}
\caption{(Color online) Temperature dependence of the full width at half maximum (FWHM) for most intense Raman modes. The vertical lines show the temperature for two charge density wave transitions.}
\end{figure}

\end{document}